\def\PRL{{ Phys. Rev. Lett.\ }\/}
\def\PRB{{ Phys. Rev. B\ }\/}
\def\PRX{{ Phys. Rev. X\ }\/}

\def\etal{{\it et.al.~}\/}

\def\be{\begin {equation}}
\def\ee{\end {equation}}
\def\ber{\begin {eqnarray}}
\def\eer{\end {eqnarray}}
\def\bers{\begin {eqnarray*}}
\def\eers{\end {eqnarray*}}

\makeatletter

\newcommand{\Rmnum}[1]{\expandafter\@slowromancap\romannumeral #1@}
\makeatother

\makeatletter
\newcommand*\env@matrix[1][*\c@MaxMatrixCols c]{%
  \hskip -\arraycolsep
  \let\@ifnextchar\new@ifnextchar
  \array{#1}}
\makeatother

\documentclass[aps,prb,showpacs,superscriptaddress,a4paper,twocolumn]{revtex4-1}
\usepackage{amsmath}
\usepackage{float}
\usepackage{color}

\usepackage[normalem]{ulem}
\usepackage{soul}
\usepackage{amssymb}

\usepackage{amsfonts}
\usepackage{amssymb}
\usepackage{graphicx}
\begin {document}

\title{Quaternary Heusler Alloy: An Ideal Platform to Realize Triple Point Fermion}

\author{C. K. Barman}
\thanks{These two authors have contributed equally to this work}
\affiliation{Department of Physics, Indian Institute of Technology, Bombay, Powai, Mumbai 400 076, India}

\author{Chiranjit Mondal}
\thanks{These two authors have contributed equally to this work}
\author{Biswarup Pathak}

\affiliation{Discipline of Metallurgy Engineering and Materials Science, IIT Indore, Simrol, Indore 453552, India}

\author{Aftab Alam}
\email{aftab@iitb.ac.in}
\affiliation{Department of Physics, Indian Institute of Technology, Bombay, Powai, Mumbai 400 076, India}
\date{\today}

\begin{abstract}
The existence of three fold rotational, mirror and time reversal symmetries often give rise to the triply degenerate nodal point (TP) in the band structure of a material. Based on point group symmetry analysis and first principle electronic structure, we predict, in this article, a series of quaternary Heusler alloys host an ideal platform for the occurrence of TP. We simulated, the projection of these TPs onto the (111) and (100) surfaces lead to form topological Fermi arcs, which may further be detected by scanning tunneling spectroscopy and angle resolved photoemission spectroscopy. These Fermi arcs arise due to the symmetry protected band degeneracies, which are robust and can not be avoided due to the non-trivial band topology. Interestingly the TPs, in these class of Heusler alloys are far away from the $\Gamma$ point along C$_3$ axes, which allow to overcome the experimental difficulties over previously studied hexagonal and HgTe-type compounds. 

\end{abstract}
	
\maketitle

\par {\it \bf Introduction}:\ Nontrivial Fermi surface topology in metals and semimetals has recently been more exciting research frontiers over gaped systems. Unlike topological insulators, topological metals and semimetals\citep{ZHCK2010,XQSCZ2011,JEMOORE,TDAS2016} are interesting not only because of their rich surface physics but also for the exotic nature of linear band crossing nature in bulk. The semimetals have  further been classified into Dirac semimetal (DSM),\cite{DSM2012,BOHM2014,Na3Bi2014,Na3Bi2015,Cd3As22014,CdAs2014,AMgBi2017,MAl32018} Weyl Semimetal (WSM),\cite{RMP2018,PENGLI2017,AASNature2015,SYXU2015,Xiangang2011,BSINGH2012,
HWENG2015,LiuVanderbilt2014} and Nodal-line semimetal (NLS)\cite{RuiYu2015,Rappe2015,GuangBian2016,GTR2016,JTWang2016,Huaqing2017,Ca3P2017,LLW2018} depending on the dimensionality of the band crossing points  and Fermi surface topology in the momentum space.The Fermi surface of DSM and WSM have been experimentally verified to be zero-dimensional discrete points in the bulk Brillouin zone (BZ), where as, it is one-dimensional close loop for NLS. The quasi-particle counter part to elementary Dirac and Weyl fermion in the standard model have been mimicked near the four fold Dirac nodes and two fold Weyl nodes in their low energy excitations. In contrast, NLS does not have a direct analogue to the elementary quasi particle in the high energy physics. 

Another example of quasi-particle excitation in condensed matter physics which does not have elementary particle counter part in the quantum field theory is topological triple point semimetal (TPSM).\cite{ZZhu2016,NexusFermion,HWeng2016,InAsSb2016,XZhang2017,GangLi2017,Jianfeng2017,PengGuo2018} TPSM is essentially protected by crystal point group symmetry and believed to be same kind of an intermediate state between four fold Dirac fermion and two fold Weyl fermion system. For example, having three fold rotational symmetry with a vertical mirror plane in a non-magnetic crystal can lead to a pair of triply degenerate nodal points along C$_3$ axis in the absence of crystal inversion. A Dirac node can be achieved by keeping the inversion intact in such an symmetry environment. If vertical mirror symmetry is further broken, it produces four Weyl nodes from two triple points reducing the band degeneracies. 

Although it is theoretically quite straightforward to achieve TPSM states but realizing such states near the Fermi level is quite fragile. Moreover, stabilizing the nodes are too hard as they need fine tuning of chemical composition. Three component fermion which is prohibited in the presence of Lorentz invariance have been predicted in both symmorphic and non-symmorphic space group symmetry.\cite{ZZhu2016} However, the hunt for a suitable material class is still ongoing for table top experiment; such as, observation of Fermi arcs using surface sensitive photo emission spectroscopy. Some of the materials such as WC,\cite{ZZhu2016} ZrTe,\cite{ZrTe2016} MoP,\cite{MoP2017} MgTa$_{2-x}$Nb$_x$N$_3$,\cite{MgTa2N32018,pureMgTa2N3} strained-HgTe\cite{HgTe2013} which have been predicted to be three component Fermion system belongs to hexagonal space group. For such a space group material, (001) surface is the most natural choice to cleave. As the TPs are protected by C$_{3v}$ point group symmetry, they lie on the C$_3$ axis. Therefore, projecting the TPs on (001) facet of hexagonal compound results in disappearance of the TP induced Fermi arcs. This scenario has been observed in case of MoP.\cite{MoP2017} Very recently, a bunch of ternary half-Heusler compounds\cite{YanBinghai2017,HfIrAs2018} have been theoretically predicted to host TPSM state along $\Gamma$ - L direction in BZ. As the little group of $\Gamma$ for these compounds is T$_d$, they contain four independent C$_3$ axis. Projecting the TPs on (111) surface will produce six Fermi arcs that connect the projected triple point nodes as described in Ref.[\cite{YanBinghai2017,HfIrAs2018}]. This class of material readily solve the previously described problem on hexagonal systems. Unfortunately, the position of TPs on the C$_3$ axis in some of these compounds\cite{HfIrAs2018,HgTe2013,YanBinghai2017} are extremely close to $\Gamma$ point. This results in extremely small Fermi arcs on its (111) facets, identification of which is a major constraint in experimental measurements. 

Here, we introduce another class of Heusler alloy which is quaternary and produce a triple point node along the $\Gamma$ - L direction in BZ. In this class of quaternary Heusler alloy, the three fold degenerate nodal point occur far away from $\Gamma$ point which consequently yields a decent sized Fermi arcs. This, in turn immediately removes the experimental constraints over some of the previous studies in HgTe\cite{HgTe2013} and ternary Heusler alloys.\cite{HfIrAs2018,YanBinghai2017}

\par {\it \bf Computational Details}:\
First principle calculations were performed using Vienna Ab-initio Simulation Package (VASP)\cite{GKRESSE1993,JOUBERT1999} based on density functional theory (DFT). To describe the exchange and correlation functional, we adopt generalized gradient approximation by Perdew-Burke-Ernzerhof (PBE).\cite{JOUBERT1999} Plane wave basis set using Projector Augmented Wave (PAW)\cite{PEBLOCHL1994} method was used with an energy cut off 500 eV. Total energy (force) was converged upto 10$^{-6}$ eV/cell (0.001 eV/\r{A}). Brillouin zone(BZ) integrations were performed using 12$\times$12$\times$12 $\Gamma$-centered k-mesh. The spin-orbital coupling (SOC) effect was included in all the calculations. A tight binding model was constructed using maximally localized wannier functions (MLWF)\cite{MARZARI1997,SOUZA2001,VANDERBIT} to closely reproduce the bulk band structure. The iterative Greens function\cite{DHLEE_I,DHLEE_II,SANCHO1985} method implemented in Wannier$\_$tool package\cite{QuanSheng}  was employed to calculate the Fermi arcs and surface dispersions. 

\begin{figure}[t]
\centering
\includegraphics[width=\linewidth]{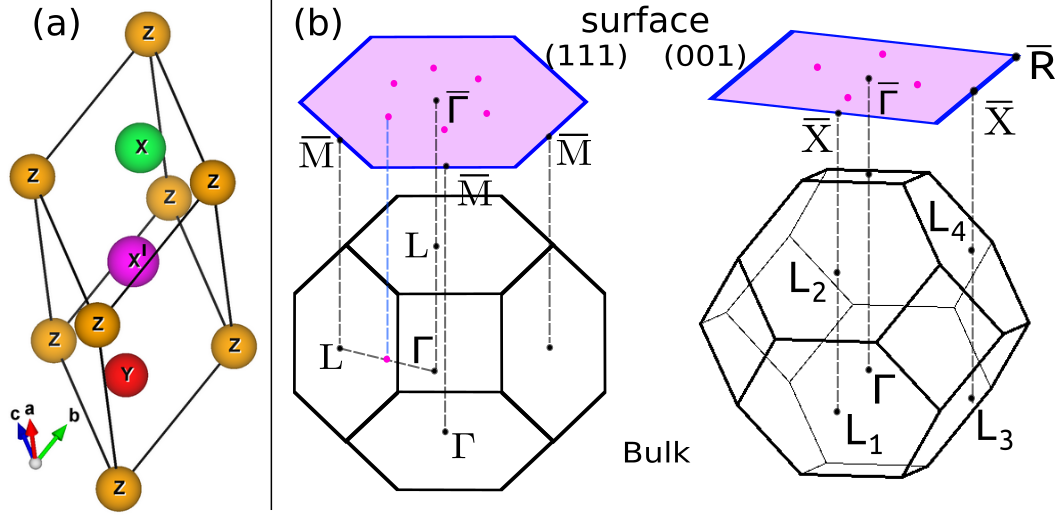}
\caption{(Color online) (a) Primitive crystal structure of quaternary Heusler alloy XX$^{'}$YZ, (b) pictorial depiction of bulk and surface BZ with high symmetry points. The projection of L points on both (111) and (001) surfaces are shown by dotted lines. Magenta dots are projection of TPs on surface.  }
\label{fig1}
\end{figure}


{\par} {\it \bf Results and Discussion:} Here, in this study, we present a bunch of quaternary half-Heusler alloy, which provide an ideal platform (as the TPs are located far away from zone center and it readily removes the bottleneck of cleaved surface in hexagonal system) to realize triple point semimetal states from both experimental and theoretical front. The crystal structure and Brillouin zone (BZ) of quaternary Heusler alloy (space group F$\bar{4}$3m\cite{Claudia2011}) is described in Fig.~\ref{fig1}. The primitive cell possesses four atoms X, X$^{'}$, Y and Z at  wyckoff positions 4c, 4d, 4b and 4a respectively. Figure~\ref{fig1}(b,c) shows the BZ of bulk cell and the projected BZ for (111) and (001) surface. The magenta dots on surface BZ will be discussed later.

\begin{figure}[bt]
\centering
\includegraphics[width=\linewidth]{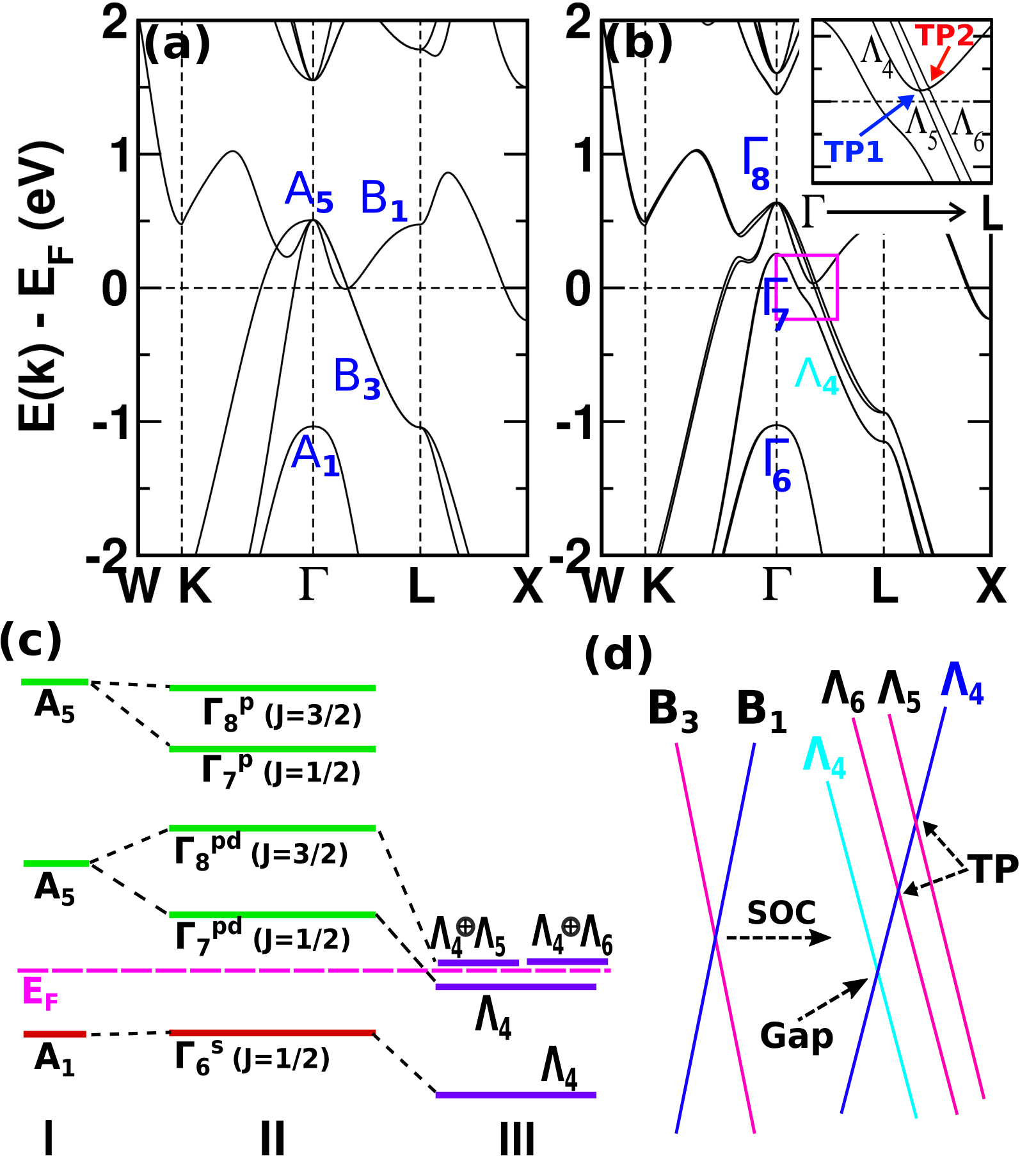}\\
\caption{(Color online) Bulk band structure of LiMgPdSb (a) w/o-soc and (b) w/-soc. The inset in (b) shows the enlarged picture around the triple point crossing region marked by colored box. (c) The schematic representation of band evolution near the Fermi level. Region I represents the combined effect of crystal field splitting and orbital hybridization in the vicinity of $\Gamma$ point. Region II represents the effect of SOC on band splitting (described in text). Region III is in the vicinity of TP where $\Gamma^{pd}_8$ bands transformed into degenerate $\Lambda_{4,5,6}$ at E$_F$ and  $\Gamma^{pd}_7$ appear below E$_F$ with $\Lambda_{4}$ character. (d) Schematic illustration for generation of triply degenerate nodal point for non-centro-symmetric crystal with C$_{3v}$ point group. $\Lambda_i$ (B$_i$) are the irreducible representations of C$_{3v}$ with(out) SOC effect. The triple points are denoted by TP. Inclusion of SOC generates TPs at the crossing of $\Lambda_{5,6}$ and $\Lambda_4$. The band crossing between two same IRs ($\Lambda_4$) opens up a gap as shown by intersection point of indigo and blue color line.}
 
\label{fig2}
\end{figure}

{\par} We take LiMgPdSb as a prototype system and provide a detailed calculation on it. The electronic structure of LiMgPdSb compound without SOC is shown in Fig.~\ref{fig2}(a). The crystal structure of cubic Heusler alloy\cite{Claudia2011} with space group 216 possesses the T$_d$ point group symmetry. To get better insight about the non-triviality and the formation of TP, we describe band symmetries based on point group analysis. In Fig.~\ref{fig2}(a), due to the cubic tetrahedral crystal field splitting, the valence band maxima (VBM) at $\Gamma$ point hold three fold degenerate A$_5$ representation which is composed of $(p,d)-$orbitals. While the second highest occupied s-like band posses A$_1$ representation of T$_d$ point group. Along $\Gamma$-L direction, the bands have B$_1$ \& B$_3$ representation under C$_{3v}$ point group, which contain four independent C$_3$ axis along $\Gamma$-L direction with three vertical mirror plane ($\sigma_{v}$) of symmetry.

{\par} In the presence of SOC , the three fold degenerate A$_5 $ states at $\Gamma$ point split into four fold $\Gamma_8$ (J$_z$= $\pm$ $\frac{3}{2}$) states  and two fold degenerate $\Gamma_7$ (J$_z$= $\pm$ $\frac{1}{2}$) states. Similarly, the A$_1$ state transform into $\Gamma_6$ (J$_z$= $\pm$ $\frac{1}{2}$). $\Gamma_6$, $\Gamma_7$ and $\Gamma_8$ are the representation of T$_d$ considering the 2$\pi$ rotation of spin subspace. Now, the s-like $\Gamma_6$ state stays, in energy, below the p-like $\Gamma_8$ states, resulting an inverted band order between $\Gamma_6$ and $\Gamma_8$ states. Such non-trivial band topology serves as to conduct unique features on surface dispersion, which will be discussed in details later in the manuscript. 

Figure~\ref{fig2}(c) describes the the origin of triple point from the perspective of chemical hybridization and crystal field splitting of atomic levels. The Pd atom is located at the tetrahedral position in crystal structure.  Without SOC, in the vicinity of $\Gamma$ point (in region I), $t_{2g}$($d_{xy}$,$d_{yz}$,$d_{zx}$) orbitals hybridized with Sb(p)-orbitals and form A$_5$ level near E$_F$. The $e_g$($d_{x^2-y^2}$, $d_{z^2}$)-orbitals (not shown here) stays far below E$_F$. On the other-hand, the high energy lying three fold degenerate A$_5$ states are contributed by Pd(p)-orbitals. In addition, the A$_1$ state below to that A$_5$ level are composed of s-like orbitals of (Pd,Mg,Sb) atoms. In region-II at/near $\Gamma$-point, the influence of SOC splits the hybridized A$_5$ state into J=$\frac{1}{2}$ and J=$\frac{3}{2}$ states. Along $\Gamma$-L direction, the $\Gamma^{pd}_8$ splits into $\Lambda_4$, $\Lambda_5$, and $\Lambda_6$ states at E$_F$ in the vicinity of triple point as shown in region III, Fig.~\ref{fig2}(c). $\Gamma^{pd}_7$, on the other hand, goes below to E$_F$, represented as two-dimensional irreducible representation (IR) $\Lambda_4$. 

{\par} Now we will discuss the general formation mechanism of triply degenerate nodal points in band structure with the help of point group symmetry analysis. A system having little group of C$_{3v}$ along some high symmetry direction, whose elements are thee fold rotation (C$_3$) and three $\sigma_v$ mirror, provides two one-dimensional IRs ($\Lambda_5$ \& $\Lambda_6$) and one two-dimensional IR ($\Lambda_4$) in its double group representation. In general, the effective mass of $\Lambda_5$/$\Lambda_6$ and $\Lambda_4$ are opposite.\cite{YanBinghai2017} So, there could be some accidental degeneracy between $\Lambda_5$/$\Lambda_6$ and $\Lambda_4$ when they disperse over the BZ along C$_{3v}$ axis. Consider a system which does not have inversion center but preserve the time reversal symmetry, the Kramer's degeneracy will be broken for $\Lambda_4$ and $\Lambda_5$ over the BZ except at time reversal invariant momenta (TRIM) points. In such a situation the dimension of degeneracy between $\Lambda_5$( or $\Lambda_6$) and $\Lambda_4$ are three fold at their crossing points. Hence they form triply degenerate nodal points when they cross each other. A cartoon diagram of formation of TP explaining above mechanism, is shown in the Fig.~\ref{fig2}(d). The existence of such TP near the Fermi level strongly influence the low energy excitation which in turn provide abnormal transport anomaly, non-trivial Fermi arc topology etc. Interestingly, the above prescription is valid for both symmorphic and non-symmorphic space group.\cite{PengGuo2018}
  
{\par} Let us now try to understand the realization of triple points in our system LiMgPdSb based on the general mechanism described above. If we closely look at the evolution of bands from w/o-SOC to w/-SOC (i.e. Fig.~\ref{fig2}(a) and \ref{fig2}(b)), the transformation of bands (along C$_{3v}$ axes) from simple group to double group follows $-$ B$_1$ $\rightarrow$ $\Lambda_4$ and B$_3$ $\rightarrow$ $\Lambda_4 \oplus \Lambda_5 \oplus \Lambda_6$. i.e, B$_1$ bands transform to $\Lambda_4$ bands, where as B$_3$ bands split into two singly degenerate $\Lambda _5$ \& $\Lambda _6$ bands and one doubly degenerate $\Lambda_4$ bands. Fig.~\ref{fig2}(b)shows that near Fermi energy $\Lambda_4$, $\Lambda_5$ and $\Lambda_6$ band almost linear with same slope upto a large momenta about 0.34\textbf{\textit{k}} along $\Gamma$-L direction. Beyond that, the $\Lambda_5$ and $\Lambda_6$ still disperse linearly towards low energy but $\Lambda_4$ changes its mass and disperse almost quadratically towards higher energy and intersect both $\Lambda_5$ \& $\Lambda_6$. The reason behind these kind of sudden change of slope by $\Lambda_4$ band could be due to the band repulsion of another $\Lambda_4$ band which originates from the p-like $\Gamma_7$(J=1/2) band at $\Gamma$ point. The intersection points of $\Lambda_4$ with $\Lambda_5$ and $\Lambda_6$ are the triply degenerate nodal points TP1 and TP2 with the position in momentum space to be (0.170, 0.170, $\pm$0.170)$\frac{\pi}{c}$ and (0.183, 0.183, $\pm$0.183)$\frac{\pi}{c}$ respectively.

\begin{figure*}[t]
\centering
\includegraphics[width=\linewidth]{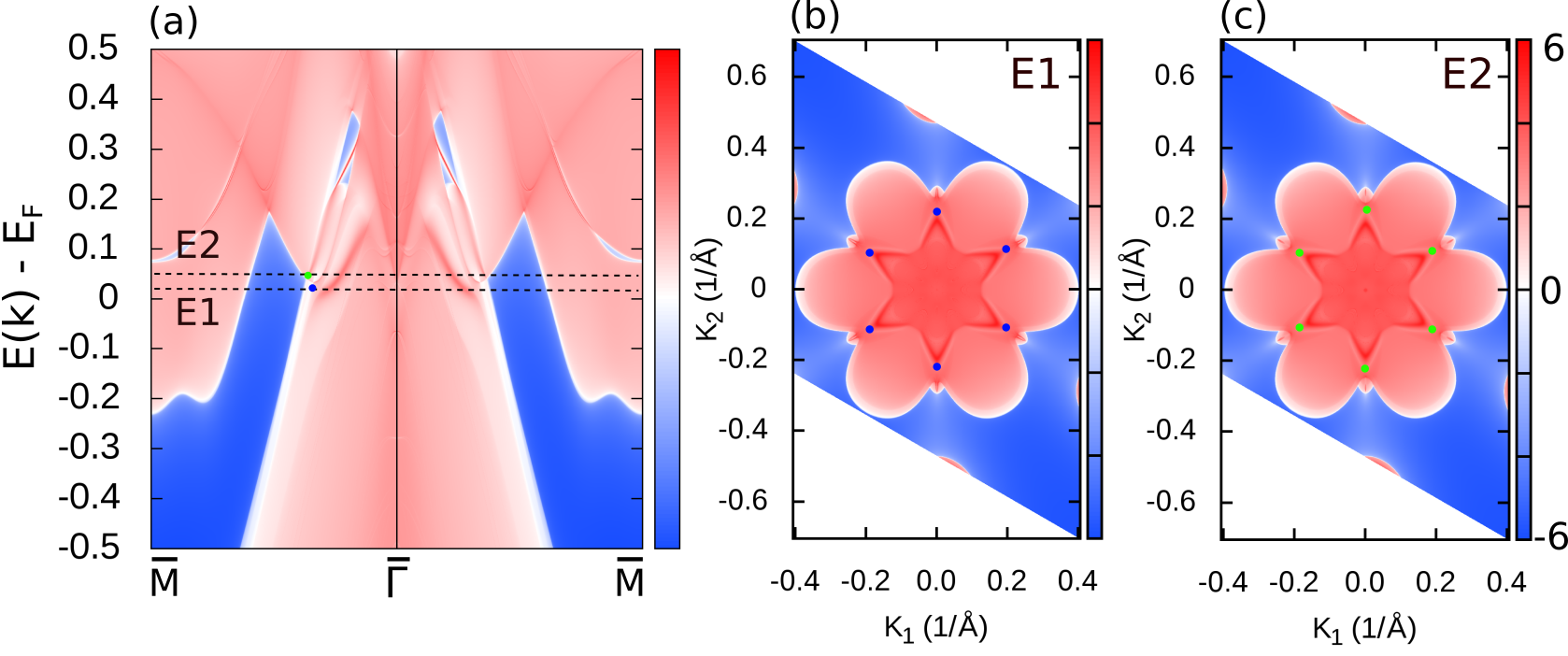}
\includegraphics[width=\linewidth]{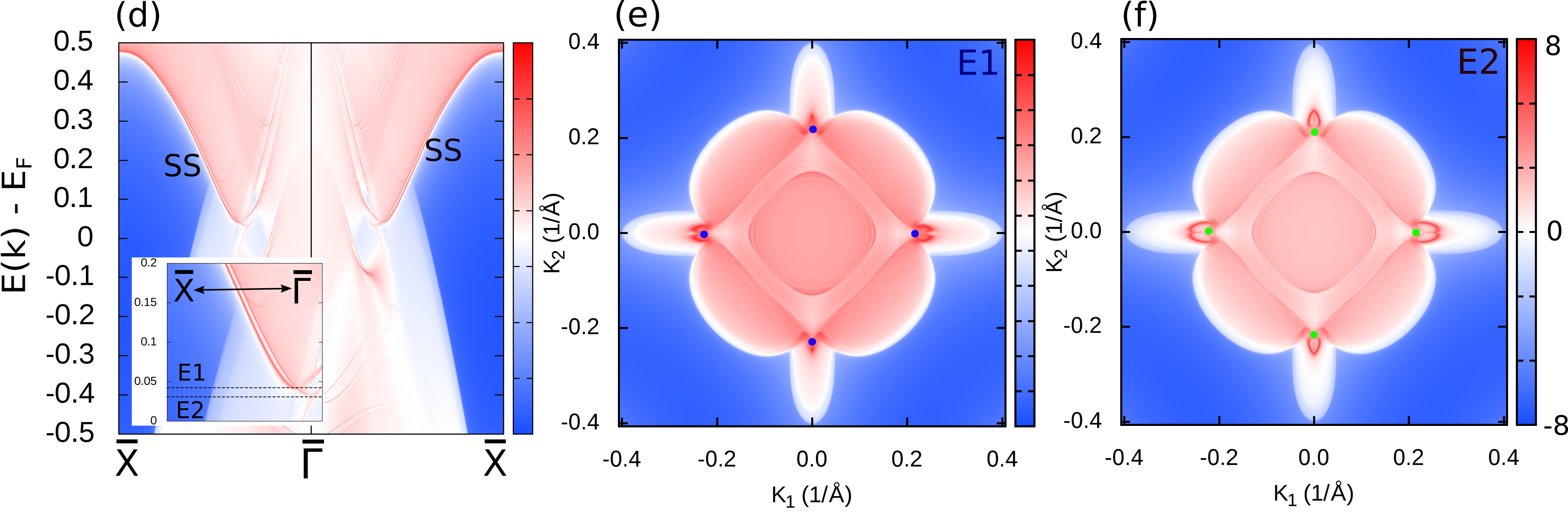}
\caption{(Color online) Surface density of states (a,d) and Fermi arcs (b,c and e,f) for (111) surface (top panel) and (001) surface (bottom panel). The constant energy map at E1(E2) is for TP1(TP2). The triple points are marked by blue and green dots for two TPs. }
\label{fig3}
\end{figure*}


\begin{table}[t]
\begin{ruledtabular}
\caption{ The compounds name, their optimized lattice parameter, number of TPs with their position in terms of energy position($\Delta\epsilon$) with respect to E$_F$ and distance ($\Delta_k$) in \% (along $\Gamma$-L line) from $\Gamma$ point along C$_3$ axes. The energy position of TP1 for each material is given, while the energy position of TP2 is very close to that of TP1.  }
\label{Table1}
\begin{tabular}{c c c c c c c}
 Compound &  a$_{opt}$ (\r{A}) &  \#  & $\Delta_k$  & $\Delta\epsilon$ (eV) \\
\hline
LiMgPdSb &  6.55 & TP1 & 34.07\% &     0.03                \\
               & & TP2 & 36.69\% &                          \\
LiMgPtSb &  6.53 & TP1 & 47.70\% &  $-$0.36                  \\
               & & TP2 & 56.49\% &                            \\
LiMgAuSn &  6.69 & TP1 & 42.36\% &	   0.13	                   \\
               & & TP2 & 46.29\% &                              \\			 
LiMgPdSn &  6.51 & TP1 & 23.84\% &	   1.72                      \\
               & & TP2 & 25.62\% &                                \\
LiMgPtSn &  6.49 & TP1 & 37.68\% &     1.54		                   \\
               & & TP2 & 42.25\% &               
\end{tabular}
\end{ruledtabular}
\end{table}

{\par} Now, given four C$_3$ axes ($\Gamma$-L line) in the bulk BZ, eight pair of TPs form in the first BZ. Projection of three pair of these TPs onto the (111) surface form hexagon surrounding the $\bar{\Gamma}$ point in surface BZ, while the fourth pair of TP exactly projects on the $\bar{\Gamma}$ point. The projection of these six TPs around the $\bar{\Gamma}$ point are pictorially depicted by magenta dots on (111) surface in Fig.~\ref{fig1}(b). The TP2(TP1) in LiMgPdSb compound is around $\sim$36.69\%(34.07\%) of $\Gamma$-L distance away from $\Gamma$ point which is large enough for experimental observations. The position of these TPs along energy axis also extremely important to be probed from a suitable experiment (photoemission spectroscopy vs. scanning tunneling microscopy). In the present system, TP1 lies extremely close to the Fermi level (E$_F$); $\sim$0.03 eV above E$_F$. TP2 position is close to TP1.

\begin{figure}[b]
\centering
\includegraphics[width=\linewidth]{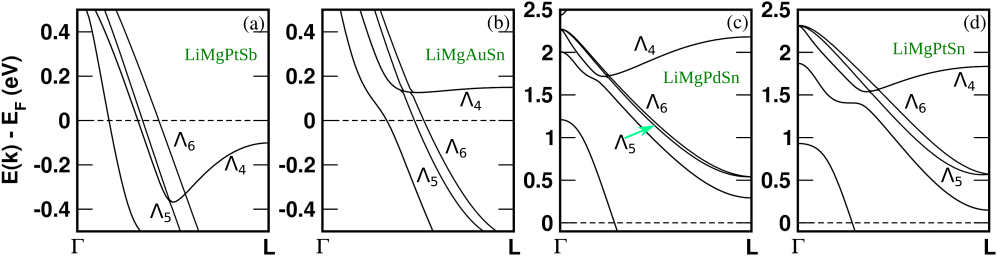}

\caption{(Color online) Bulk band structure of (a) LiMgPtSb, (b) LiMgAuSn, (c) LiMgPdSn and (d) LiMgPtSn with spin-orbit coupling. The pair of triple points are shown at the crossing of $\Lambda_4$ and $\Lambda_{5,6}$  IRs along $\Gamma$-L direction. }
\label{fig4}
\end{figure}

{\par}A similar observation of TPs are also reported on strained HgTe and ternary half-Heusler compounds.\cite{HgTe2013,YanBinghai2017,HfIrAs2018} But unfortunately TPs in these materials are close to the $\Gamma$ point. There exists few other compounds\cite{YanBinghai2017} where the position of TPs lie at large momenta due to the "peculiar double-valley" type dispersion of $\Lambda_4$ bands. The mechanism for the formation of triple point nodes in these compounds are quite different compared to our materials. In our case, the emergence of large momenta TPs are in its simplest form and are expected to be more interesting for experimental probing.

Similar to the WSM, TPSMs also hold the signature of bulk band degeneracies onto its surfaces. Hence surface dispersion and Fermi arcs of TPSM are worthy of careful investigation both from theoretical and experimental front. Since a TP is equivalent to superposition of two degenerate Weyl points, a natural expectation is the appearance of two Fermi arc from a typical TP. These two Fermi arcs, originated from a TP, are further expected to connect with two other nearest TPs in the Surface BZ. In Fig.~\ref{fig3}, we present the surface density of states and the Fermi arcs of (111) and (001) surface. Energy scales of TPs on surfaces are indicated by dashed reference lines (E1 \& E2). For (111) surface in Fig.~\ref{fig3}(a), a pair of surface state originates from two TPs and disperse along $\bar{\Gamma}$ point and they are supposed to be degenerate at the zone center as $\bar{\Gamma}$ is one of the time reversal point of the (111) surface BZ. But they get masked by the bulk bands and disappear near $\bar{\Gamma}$ as they propagate from zone boundary to zone center. However, the degeneracy of these two TP induced surface states are clear at $\bar{X}$ in (001) surface spectrum as shown in Fig.~\ref{fig3}(d). Here, a pair of surface state (SS) emerges from two TPs and disperse towards $\bar{X}$ (as shown in inset of Fig.~\ref{fig3}(d)) and become degenerate at $\bar{X}$ as it is one of the four TRIM point of (001) surface BZ. Fig.~\ref{fig3}(b,c) and~\ref{fig3}(e,f)  show the constant  energy contour map at energy E1=0.033 eV and  E2=0.041 eV corresponding to TP1 and TP2 respectively. As shown in Fig.~\ref{fig3}(b-c), the Fermi arcs  appear from each six TPs projected on (111) surface and they merged to the nearest TPs to form a hexagon shape surrounding the $\bar{\Gamma}$ point. The nature of Fermi arcs are further preserved by C$_3$ symmetry in (111) surface. However, for (001) surface four TPs are projected on surface BZ. The projection of TPs are also schematically shown in Fig.~\ref{fig1}(b). These four projected TPs on (001) surface are connected by Fermi arcs which is different in shape compare to (111) surface Fermi arcs, as illustrated in Figure~\ref{fig3}(e-f). 
{\par} In addition to LiMgPdSb, we have investigated several other compounds in the quaternary Heusler class and found four of them quite interesting. These compounds indeed show similar TPs, but at much different momenta distance and both the $+$ve and $-$ve side of the E$_F$. These compounds are LiMgPtSb, LiMgAuSn, LiMgPdSn, and LiMgPtSn, whose SOC band structures are shown in Fig.~\ref{fig4}. The optimized lattice parameter, and the location of TPs in terms of energy position ($\Delta\epsilon$) with respect to E$_F$ and momenta distance $\Delta_k$ along $\Gamma$-L are shown in Table~\ref{Table1}. These TPs being away from the $\Gamma$-point, should be easier to probe and hence awaits experimental confirmation.


\par {\it \bf Conclusion}:\ In summary, we show the occurrence of triple point Fermionic state in quaternary Heusler alloys. The symmetry protected band degeneracies are the recipe of TP formation. We found, along $\Gamma$-L direction, two bands from two different IRs ($\Lambda_4$ and $\Lambda_{5,6}$) of C$_{3v}$ point group crosses each others and hence form triple point. The surface states on (111) and (100) miller plane are investigated and shown. The projected TPs on (111) surface revealed a hexagonal shape Fermi arcs which is a band topology mediated phenomena arises due to the non-trivial band degeneracies. The shape of Fermi arcs depends on the type of cleaved surface and hence the number of TPs projected on the surface. The observed TPs in our system of interest are far away from the Brillouin zone center along C$_3$ axes. Thus, we believe that quaternary Heusler alloy provide an ideal platform for triple point Fermionic state and demands strong experimental investigation using photoemission spectroscopy and tunneling microscopy.

C.K.B and C.M acknowledge MHRD-India for financial support.


\newpage


\end{document}